\documentclass{WileyChemistry-template}
\usepackage{bm}
\usepackage{graphicx}
\usepackage{subcaption} %

\title{\raggedright Computational and Experimental Investigation of Chiral and Achiral 2D Organic Lead Bromide Perovskites: Octahedral Distortions and Electronic and Optical Properties}

\author{
\begin{minipage}{\textwidth}
	Md Mehdi Masud,\textsuperscript{[a]}
    Jarek Viera,\textsuperscript{[b]}
    Azza Ben-Akacha,\textsuperscript{[b]}
    Biwu Ma,*\textsuperscript{[b]}
    David A. Strubbe*\textsuperscript{[a]}
\end{minipage}
}

\newcommand{\affiliation}{
\begin{itemize}

\item[{[a]}] Md Mehdi Masud, David A. Strubbe\\
Department of Physics\\
University of California, Merced\\ Merced, California 95343, USA\\
E-mail: %
dstrubbe@ucmerced.edu

\item[{[b]}] Jarek Viera, Azza Ben-Akacha, Biwu Ma\\
Department of Chemistry and Biochemistry\\
Florida State University\\
Tallahassee, Florida 32306,
USA\\
E-mail: bma@fsu.edu\\
\end{itemize}
}

\renewcommand{\dedication}{
	\begin{minipage}{\textwidth}
		
	\end{minipage}
}

\renewcommand{\abstract}
{We present a computational investigation, in conjunction with synthesis and experimental characterization, into the structural, electronic, and optical properties of layered 2D organic lead bromide perovskites. We contrast materials based on the chiral (R/S)-4-fluoro-$\alpha$-methylbenzylammonium (R/S-FMBA), which have been shown to lead to bright room-temperature circularly polarized luminescence, with the similar achiral 4-fluorobenzylammonium (FBA). Using density functional theory (DFT) with van der Waals (vdW) corrections, we study relaxed structures (compared with X-ray diffraction, XRD) and optical absorption spectra (compared with experiments), as well as bandstructure and orbital character of transitions. We develop and provide a Python code to calculate octahedral distortions and compare DFT and XRD results, finding that vdW corrections are important for accuracy and that DFT overestimates octahedral tilt angles. (FMBA)\textsubscript{2}PbBr\textsubscript{4} shows among the largest tilt angle differences (often termed $\Delta \beta$) reported, $14^\circ$, indicating strong inversion symmetry-breaking which enables its chiral emission. The lowest-energy optical transitions involve the perovskite only and are polarized within the layer. This work furthers understanding of structure-property relations with applications to optoelectronics and spintronics.}

\begin{document}

\twocolumn[\vspace{-1.5cm}\maketitle\vspace{-1cm}
	\textit{\dedication}\vspace{0.4cm}]
\small{\begin{shaded}
		\noindent\abstract
	\end{shaded}
}

\begin{figure} [!b]
\begin{minipage}[t]{\columnwidth}{\rule{\columnwidth}{1pt}\footnotesize{\textsf{\affiliation}}}\end{minipage}
\end{figure}

\section*{Introduction}
\label{introduction}
\sloppy
Two-dimensional (2D) organic-inorganic hybrid perovskites (OIHPs) \cite{Miura2022,Lin2022,Liu2024} have garnered significant attention due to their unique optoelectronic properties, structural versatility, and potential for chiroptoelectronic and spintronic applications \cite{Liu2024,Duan2024,Kucheriv2024}. These materials are characterized by their ability to incorporate chiral organic ligands \cite{Ou2024}, which impart chirality to the inorganic sublattice, enabling applications in circularly polarized light detection, optical information processing, and spin-selective devices. Of particular interest are the chiral 2D OIHPs such as (R)- and (S)-4-fluoro-$\alpha$-methylbenzylammonium (FMBA)-based lead bromide perovskites, (R/S-FMBA)\textsubscript{2}PbBr\textsubscript{4} \cite{Zhao2023}. These materials have demonstrated promising properties, including room-temperature circularly polarized luminescence and enhanced quantum yields when crystallized into oriented films, as shown by recent experimental works \cite{Zhao2023}.

Despite advances in synthesizing and characterizing these materials, understanding of their structural and electronic properties at the atomic level remains limited. Experimentally observed phenomena such as distortions in [PbBr\textsubscript{6}]\textsuperscript{4-} octahedra, hydrogen bonding effects, and their influence on electronic band structure and optical transitions are still not fully understood. The importance of octahedral distortions in perovskites has been appreciated for some time \cite{zhou2005universal,Woodward1997}.
Studies in organic-inorganic perovskites have shown that larger tilt angles between octahedra usually indicate more exciton trapping and wider bandgap \cite{prasanna2017band,shao2022unlocking}. Out-of-plane distortions have been established as a descriptor in 2D perovskites that is correlated with broadband emission \cite{smith2017structural}. More recently, the deviation among tilt angles of different pairs of octahedra has been identified as a metric of inversion symmetry breaking in chiral perovskites, correlated with Rashba/Dresselhaus spin splittings \cite{Jana}. Specifically for FMBA materials, the previous studies \cite{Zhao2023} highlighted significant out-of-plane distortions in the equatorial plane of the inorganic octahedra and studied distortions within octahedra, but did not fully connect structural distortions to their electronic and optical behaviors.

To gain further insight into structure-property relationships, we synthesized (R/S-FMBA)\textsubscript{2}PbBr\textsubscript{4} materials experimentally, as well as a related achiral reference material 4-fluorobenzylammonium (FBA)$_2$PbBr$_4$. We then carried out a comprehensive computational study, using density functional theory (DFT) with van der Waals corrections to provide an in-depth understanding of the behavior of these materials. We investigated structural relaxation and octahedral distortion parameters, with comparison to the X-ray diffraction (XRD) structure. We studied electronic bandstructure, atomic orbital contributions, and polarized optical absorption spectra, compared with measured UV/Vis absorption spectra.
We developed a Python code to easily calculate octahedral distortion parameters, adapting previous formulae \cite{smith2017structural,zhou2005universal,Woodward1997,Freitas2025} and the MATLAB code from Ref. \cite{smith2017structural} which calculates in-plane and out-of-plane distortion angles.
Other existing software includes Octadist \cite{Ketkaew2021} and the visualization code VESTA \cite{momma2011vesta} which can both calculate some octahedral distortion parameters.  
Our code brings together the calculation of a variety of intra- and inter-octahedral distortion parameters in a single tool, and works in Python which is free, widely available, and increasingly commonly used.
This tool, included in the Supplementary Information, enables the quantification of bond length deviations and bond angle distortions, for understanding structure-property relationships in perovskite materials for optoelectronic and spintronics applications.

\section{Methods}
To investigate the structural, electronic, and optical properties of layered 2D lead bromide perovskites, we experimentally synthesized the materials and performed comprehensive density functional theory (DFT) and Random-phase approximation (RPA) calculations to compare with the experimental data.

\subsection*{Experimental Synthesis and Characterization}
\label{main_synthesis}
\begin{figure*}[ht!]
\centering
\includegraphics[scale=1.5]{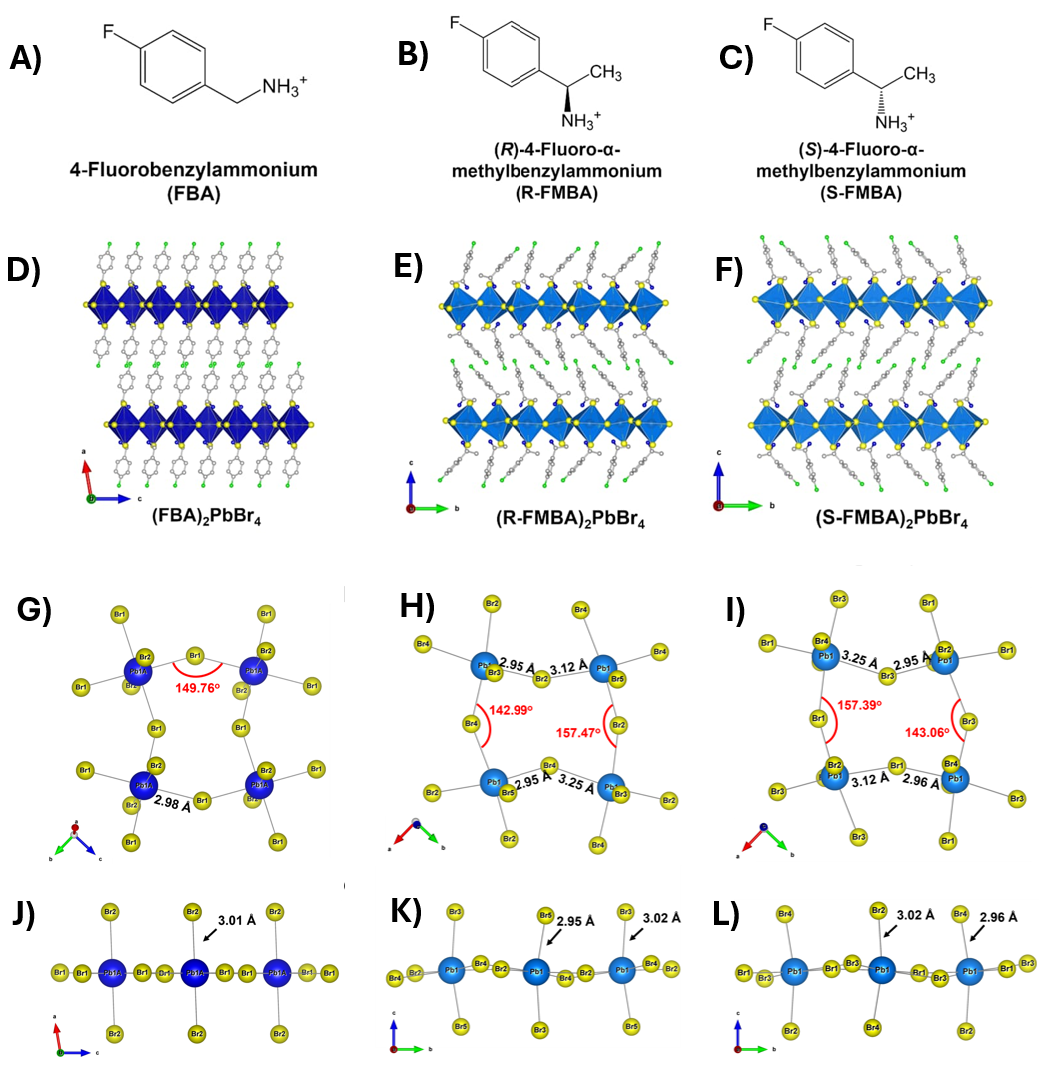}
\caption{(A--C) Organic ammonium cations, (D--F) single crystal structures of 
$(\mathrm{FBA})_2\mathrm{PbBr}_4$, $(\mathrm{R\text{-}FMBA})_2\mathrm{PbBr}_4$, 
and $(\mathrm{S\text{-}FMBA})_2\mathrm{PbBr}_4$, 
(G--I) Top-down view and (J--L) side view of the connected lead bromide octahedra 
with different bond angles and lengths from the crystal structures of 
$(\mathrm{FBA})_2\mathrm{PbBr}_4$, $(\mathrm{R\text{-}FMBA})_2\mathrm{PbBr}_4$, 
and $(\mathrm{S\text{-}FMBA})_2\mathrm{PbBr}_4$.
}
\label{fig_structure_descriptions}
\end{figure*}

\sloppy
We synthesized three layered two-dimensional (2D) perovskites: 
\((\mathrm{R\text{-}FMBA})_{2}\mathrm{PbBr}_{4}\), 
\((\mathrm{S\text{-}FMBA})_{2}\mathrm{PbBr}_{4}\), 
and \((\mathrm{FBA})_{2}\mathrm{PbBr}_{4}\). 
Briefly, we prepared these perovskites by reacting the respective hydrobromide salts 
of the organic cations with \(\mathrm{PbBr}_{2}\) in hydrobromic acid at \(100^\circ\)C.
Upon cooling, transparent plate-like crystals grew from solution. All three crystals form the same general layered-2D framework, 
composed of \(\mathrm{PbBr}_{4}^{2-}\) sheets separated by two adjacent organic cation layers, 
though the \(\mathrm{R\text{-}}\) and \(\mathrm{S\text{-}}\)forms adopt chiral 
crystal structures in the orthorhombic space group
$P2_12_12_1$, with two layers per unit cell, while the \(\mathrm{FBA}\) compound is achiral, crystallizing in the monoclinic $P2_1/c$ centrosymmetric space group with only one layer per unit cell (Fig. \ref{fig_structure_descriptions}).
The structure of (FBA)\textsubscript{2}PbBr\textsubscript{4} consists of 41 atoms in the formula unit and 82 atoms in the unit cell. In contrast, the (R/S-FMBA)\textsubscript{2}PbBr\textsubscript{4} structures each contain 47 atoms in the formula unit and 188 atoms in the unit cell. Their single-crystal X-ray diffraction (SCXRD) data confirm that all three exhibit layered-2D arrangements of corner-sharing \(\mathrm{PbBr}_6\) octahedra. 
The phase purity of the crystals and their uniformity were also
confirmed by powder XRD. Further details of the synthesis and characterization procedures can be found in the Supplementary Information.

\subsection*{Density-functional theory calculations}
Plane-wave DFT calculations were performed using Quantum ESPRESSO version 7.0 \cite{Giannozzi_2009}. The Perdew-Burke-Ernzerhof (PBE) exchange correlation-functional \cite{perdew96} was used, along with the Grimme D2 dispersion correction \cite{Grimme_2006} to account for van der Waals interactions. Norm-conserving ONCV pseudopotentials \cite{ONCV}, sourced from PseudoDojo \cite{Setten_2018}, were used to describe the interactions between core and valence electrons. These calculations were performed without the spin-orbit coupling (SOC) effect. It is often found that the Kohn-Sham bandgap estimate from a semilocal functional such as PBE is comparable to the true bandgap due to cancellation of an opening of the gap by quasiparticle effects and a closing of the gap by SOC \cite{Leppert,Filip_2024,Leppert2}.

The kinetic energy cut-off point for the wavefunction was set to 1225 eV, and a $3\times3\times1$ unshifted $k$-grid was utilized for self-consistent field (SCF) calculations. Convergence thresholds for forces and stresses were set to $10^{-4}$ Ry/bohr and 0.1 kbar, respectively. For density of states (DOS) calculations, a denser $6\times6\times2$ unshifted $k$-grid and a Gaussian 0.05 eV broadening were employed.

Variable cell relaxations were performed starting from the experimental XRD structures, leading to lattice parameters closely agreeing with the experimental values (see Table \ref{table:lattice_parameters}). Importantly, the space groups of the structures were preserved during the relaxation process. The XRD and DFT relaxed structures are provided in Supplementary Information in CIF format.

Optical absorption spectra were computed at the RPA level using the BerkeleyGW code \cite{Deslippe_2012}. In this context, RPA refers to use of the Kohn-Sham eigenvalues and wavefunctions in an independent particle approximation. These calculations employed $9\times9\times3$ $k$-point sampling with a Gaussian broadening of 0.1 eV. For the spectra, 400 occupied states and 30 unoccupied states were used for (R/S-FMBA)\textsubscript{2}PbBr\textsubscript{4}, while 188 occupied states and 30 unoccupied states were used for %
(FBA)$_2$PbBr$_4$.

\sloppy
The RPA optical absorption spectra as a function of light frequency $\omega$ were obtained via the following equation, implemented in the BerkeleyGW code \cite{Deslippe_2012}:
\begin{equation}
\epsilon_2( \omega) = \frac{16 \pi^2 e^2}{\hbar^2 \omega^2} \sum_{vck} \left| \mathbf{e} \cdot \langle v\mathbf{k} \mid \mathbf{v} \mid c\mathbf{k} \rangle \right|^2 
\delta \left( \hbar \omega - E_{c\mathbf{k}}^{\mathrm{DFT}} + E_{v\mathbf{k}}^{\mathrm{DFT}} \right),
\end{equation}
where $\textbf{e}$ is the electric-field polarization vector of the light 
and $\langle vk \mid \mathbf{v} \mid ck \rangle$ is the velocity matrix element, which quantifies the transition probability between the valence band $v$ and conduction band $c$ at a certain $k$-point. 
\( E_{ck}^{\mathrm{DFT}} \) and \( E_{vk}^{\mathrm{DFT}} \) are the DFT energy eigenvalues of the conduction band and valence bands.

The calculations were performed separately along the three Cartesian directions (\(x\), \(y\), and \(z\)), where the $z$-axis is parallel to the $c$ crystallographic direction and perpendicular to the inorganic 2D layers, to obtain the polarization-resolved \(\epsilon_2(\omega)\) data. An isotropic (unpolarized) response was subsequently obtained by averaging over the three directions:
\[
\epsilon_{2,\mathrm{unpolarized}}(\omega) = \frac{1}{3} \left[ \epsilon_{2,xx}(\omega) + \epsilon_{2,yy}(\omega) + \epsilon_{2,zz}(\omega) \right].
\]
From the dielectric function, the extinction coefficient \(k(\omega)\) was calculated as \cite{hung2022quantum}:
\[
k(\omega) = \sqrt{ \frac{ \sqrt{ \epsilon_1^2(\omega) + \epsilon_2^2(\omega)} - \epsilon_1(\omega)}{2} }.
\]
The absorption coefficient \(\alpha(\omega)\) was then determined \cite{hung2022quantum} via
\[
\alpha(\omega) = \frac{2\omega}{c} \, k(\omega),
\]
where \(c\) is the speed of light in vacuum. For comparison to experiment, spectra were plotted in terms of the wavelength $\lambda = hc/E$.

\subsection*{Determination of Octahedral Distortion Parameters}

\begin{figure*}[ht!]
\centering
\includegraphics[scale=0.9]{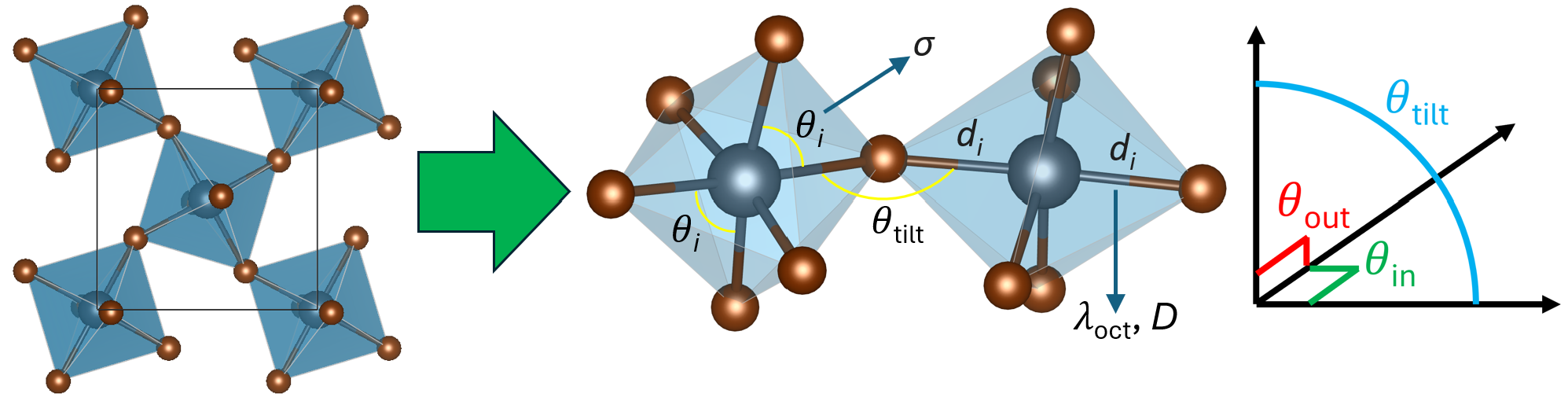}
\caption{A schematic of octahedral distortion parameters in a 2D perovskite: within an octahedron, deviations $\lambda_{\text{oct}}$ and $D$ of the bond lengths $d_i$, and deviation $\sigma$ of bond angles $\theta_i$; between neighboring octahedra, the tilt angle $\theta_{\text{tilt}}$ and its resolution into components in plane $\theta_{\text{in}}$ and out of plane $\theta_{\text{in}}$, with reference to the 2D perovskite layers \cite{smith2017structural}.}
\label{fig:octa_distortion_schematic}
\end{figure*}

\sloppy
The octahedral distortion parameters were analyzed to understand the structural variations in (FBA)$_2$PbBr$_4$ and (R/S-FMBA)$_2$PbBr$_4$. The tilting of the equatorial Pb–Br–Pb bond angles ($\theta_{\text{tilt}}$, and also referred to as $\beta$ in some literature \cite{SciAdv,Jana}) can be divided into an in-plane angle ($\theta_{\text{in}}$) and an out-of-plane angle ($\theta_{\text{out}}$) \cite{smith2017structural} (Fig. \ref{fig:octa_distortion_schematic}). To quantify the distortion of the PbBr$_6$ octahedra, three primary parameters, $D_{\text{tilt}}$, $D_{\text{in}}$, and $D_{\text{out}}$, were computed. These are defined as $D_{\text{tilt}} = 180^\circ - \theta_{\text{tilt}}$, $D_{\text{in}} = 180^\circ - \theta_{\text{in}}$, and $D_{\text{out}} = 180^\circ - \theta_{\text{out}}$. There are two angles in our chiral structures since there are inequivalent pairs of octahedra in the structure \cite{Jana}, and in some cases there can be even more distinct angles \cite{SciAdv}. The difference between these angles is defined as $\Delta D_{\text{tilt}}$ (called $\Delta \beta$ in Refs. \cite{Jana, SciAdv}). 

The bond length distortion indices were calculated as $\lambda_{\text{oct}} = \frac{1}{6} \sum_{i=1}^6 [\frac{d_i - d_0}{d_0}]^2$ and $D = \frac{1}{6} \sum_{i=1}^6 \frac{|d_i - d_0|}{d_0}$, where $d_0$ is the mean Pb–Br bond length and $d_i$ represents individual bond distances. These metrics differ only in the exponent, where $\lambda_{\text{oct}}$ is the fractional variance whereas $D$ is the fractional mean difference from the average. 
The bond angle variance, $\sigma^2 = \frac{1}{11} \sum_{i=1}^{12} (\theta_i - 90^\circ)^2$, was used to measure the deviation from ideal octahedral geometry, with $\theta_i$ denoting the Br–Pb–Br bond angles.
The calculated octahedral distortion parameters ($\lambda_{\text{oct}}$, $D$, $\sigma^2$, $D_{\text{tilt}}$, $D_{\text{out}}$, $D_{\text{in}}$)
were computed with a single Python code which provides a quantitative insight into the structural differences between the experimental (XRD) and theoretical (DFT) results for (FBA)$_2$PbBr$_4$ and (R/S-FMBA)$_2$PbBr$_4$. The code was adapted and extended from the MATLAB code provided in Ref. \cite{smith2017structural}, and carefully benchmarked against the results from that code as well as the published results.

The Python code analysis provides an efficient approach to extract these parameters systematically, which can further link to their structural, electronic and optical properties. The code is provided in the Supplementary Information.
%
The code requires atomic positions to be provided in Cartesian (Angstrom) units.

Given two octahedra centered on atoms Pb1 and Pb2, $D_{\text{tilt}}$ is calculated as the angle between the Pb1-Br and Br-Pb2 bond vectors. The code uses two distinct approaches to compute $\theta_{\text{in}}$ and $\theta_{\text{out}}$: a full 3D projection approach, as defined in Ref. \cite{smith2017structural} and used in all calculations in this paper, and a simplified 2D projection method onto the $xy$ plane.
For the 3D projection approach, $D_{\text{in}}$ is determined by projecting the Br atom onto the plane defined by Pb1, Pb2, and an atom Pb3 an adjacent octahedron, and calculating the angle within this plane. $D_{\text{out}}$ involves projecting the Br atom onto a plane orthogonal to the Pb1-Pb2-Pb3 plane. The angle between Pb1 and Pb2 within this orthogonal plane is then used to calculate $D_{\text{out}}$. The code also implements a 2D projection method, where the $xy$ plane is used as a reference instead of the plane defined by 3 Pb atoms. In practice this is done by explicitly setting $z=0$ for the Pb1, Br, and Pb2 atoms, thereby confining the geometry to the XY plane. This method is often similar to the 3D projection method, but may be somewhat different and indeed more suitable in cases of significant corrugation in the perovskite layer, in which case choice of different atoms Pb3 could give differing results.

\section{Results and Discussion}

\subsection*{Overall Structure}
\begin{figure*}[h!] \centering \includegraphics[width=0.45\textwidth]
{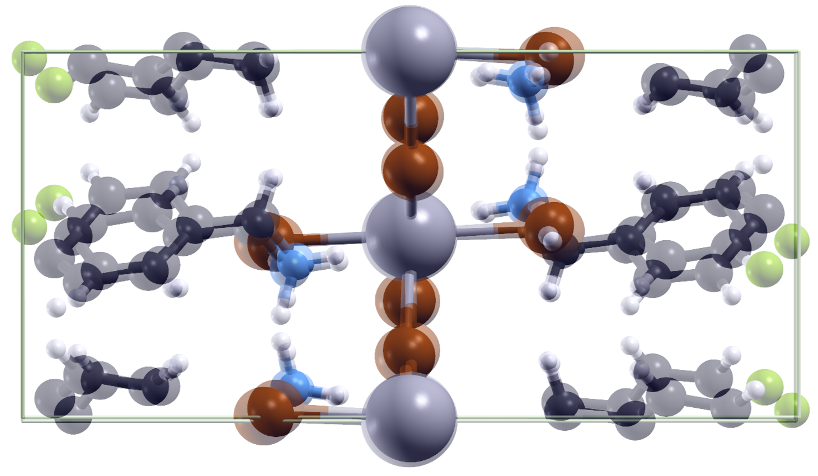} 
\includegraphics[width=0.9\textwidth]{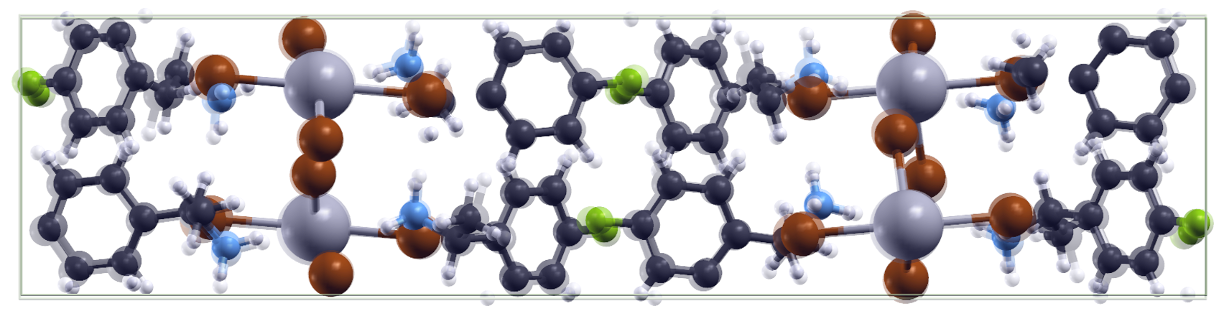}
\caption{Comparison of XRD (solid) and DFT-relaxed (partially transparent) structures of (FBA)$_2$PbBr$_4$ (top) and (R-FMBA)$_2$PbBr$_4$ (bottom).}
\label{fig:relaxed_structures}
\end{figure*}

DFT-relaxed structures were compared to structures determined by XRD to assess the accuracy of computational methods (Fig. \ref{fig:relaxed_structures}).
For (FBA)$_2$PbBr$_4$, the DFT-relaxed and XRD structures appeared similar, but the benzene rings were rotated more parallel to the $xz$ plane in the DFT calculation. This effect was much more pronounced when van der Waals corrections were not included. It is also observed that the DFT-relaxed lattice parameters are closer to the XRD structure when van der Waals corrections are used (Table \ref{table:lattice_parameters}). This motivated us to use the van der Waals corrections for further calculations on (R/S-FMBA)$_2$PbBr$_4$.
Similarly, the (R/S-FMBA)$_2$PbBr$_4$ relaxed and XRD structures were very similar except for slight rotations of the benzene ring and small displacements of other atoms.

Given the experimental synthesis, we expect the R-FMBA and S-FMBA structures to be enantiomers. The structures relaxed by DFT from the XRD coordinates were compared by reflecting the R-FMBA in the $yz$ plane and looking at differences in coordinates. The maximum displacement between the atoms in the two structures was on the order of 0.02 \AA. We ascribe any difference to small irreproducibility or noise in the synthesis and XRD, since any true difference in crystal structure is not possible by symmetry. Some further results are presented only for one enantiomer, since in this work all the properties we calculate are achiral and would be identical for exact enantiomers. The small differences in properties between R-FMBA and S-FMBA in our calculations are due only to small structural differences.

\subsection*{Lattice Parameters}
Lattice parameters from DFT were compared to those obtained experimentally by XRD (Table \ref{table:lattice_parameters}).
The relaxed lattice parameters for (FBA)$_2$PbBr$_4$ obtained with the PBE functional with Grimme-D2 van der Waals (vdW) corrections were in good agreement with the XRD parameters, particularly the angle $\beta$. Calculations without vdW corrections yielded lattice parameters that deviated further from the experimental values, confirming the importance of including dispersion corrections for layered perovskite systems.
For (R/S-FMBA)$_2$PbBr$_4$, the PBE functional with vdW corrections gave relaxed lattice parameters that were very close to those from XRD, as shown in table \ref{table:lattice_parameters}.
Based on these results, all subsequent calculations used the PBE functional with van der Waals corrections.
We also find there is close agreement between our XRD and the literature \cite{Zhao2023} (Table \ref{table:lattice_parameters}).

\begin{table*}[h!]
\centering
\caption{Relaxed lattice parameters for (FBA)$_2$PbBr$_4$ and (R/S-FMBA)$_2$PbBr$_4$ calculated using DFT with and without van der Waals corrections. Experimental lattice parameters from our synthesized materials and from literature \cite{Zhao2023} are included for comparison.}
\label{table:lattice_parameters}
\begin{tabular}{|c|c|c|c|c|c|c|c|}
\hline
\textbf{Compound} & \textbf{Method} & 
\boldmath$a$ (\AA) &
\boldmath$b$ (\AA) &
\boldmath$c$ (\AA) &
\boldmath$\alpha$ ($^\circ$) &
\boldmath$\beta$ ($^\circ$) &
\boldmath$\gamma$ ($^\circ$) \\
\hline
(FBA)$_2$PbBr$_4$ 
    & XRD [this work]            
      & 8.11 & 8.15 & 17.55 & 90.00 & 99.74 & 90.00 \\
space group: P 2$_1$/c    & DFT [PBE + vdW]            
      & 8.14 & 8.03 & 16.41 & 90.00 & 99.44 & 90.00 \\
    & DFT [PBE, no vdW]         
      & 8.30 & 8.26 & 19.07 & 90.00 & 97.94 & 90.00 \\
\hline
(R-FMBA)$_2$PbBr$_4$ 
    & XRD \cite{Zhao2023}        
      & 7.84 & 8.80 & 33.43 & 90.00 & 90.00 & 90.00 \\
space group: P 2$_1$2$_1$2$_1$    & XRD [this work]            
      & 7.88 & 8.82 & 33.78 & 90.00 & 90.00 & 90.00 \\
    & DFT [PBE + vdW]            
      & 7.76 & 8.68 & 32.48 & 90.00 & 90.00 & 90.00 \\
\hline
(S-FMBA)$_2$PbBr$_4$ 
    & XRD \cite{Zhao2023}        
      & 7.84 & 8.80 & 33.43 & 90.00 & 90.00 & 90.00 \\
space group: P 2$_1$2$_1$2$_1$    & XRD [this work]            
      & 7.88 & 8.82 & 33.80 & 90.00 & 90.00 & 90.00 \\
    & DFT [PBE + vdW]            
      & 7.77 & 8.68 & 32.45 & 90.00 & 90.00 & 90.00 \\
\hline
\end{tabular}
\end{table*}

\subsection*{Octahedral Distortion Parameters}
The results are presented in Table \ref{table:octahedral_distortion}. We notice reasonable agreement between our XRD and the literature \cite{Zhao2023} across the parameters, except for a notable discrepancy of the smaller $D_{\text{in}}$ value between our S-FMBA and literature (and our R-FMBA value). There are some small differences between DFT and XRD values. The distortion angles $D_{\text{tilt}}$, $D_{\text{out}}$, and $D_{\text{in}}$, as well as the $\Delta D_{\text{tilt}}$ are overestimated in DFT. We note very large values of $\Delta D_{\text{tilt}}$ for FMBA, around $14^\circ$ from the XRD structure, which are among the highest reported. For comparison, Ref. \cite{Jana} reported 3 compounds with $14-15^\circ$, and Ref. \cite{SciAdv} reported a compound with $11^\circ$. This parameter shows a very strong inversion symmetry-breaking by the octahedral distortions, revealing the cause of the bright circularly polarized emission that was observed \cite{Zhao2023}.  The values of $\Delta D_{\text{in}}$, also found in Ref. \cite{Jana} to correlated with spin splittings, are indeed even larger. By contrast, achiral FBA has $\Delta D_{\text{tilt}} = 0$.

The larger $D_{\text{out}}$ values for relaxed (R-FMBA)$_2$PbBr$_4$ and (S-FMBA)$_2$PbBr$_4$ were found to be $37.7^\circ$ and $37.8^\circ$, respectively, significantly larger than that for (FBA)$_2$PbBr$_4$ ($3.5^\circ$), indicating greater distortions induced by the larger chiral organic cations. This larger difference is also observed in XRD structures too. The values of $D_{\text{tilt}}$ and $D_{\text{in}}$ are more similar, however.

\sloppy
We further compared the octahedral distortion parameters obtained from DFT calculations with and without vdW correction. As summarized in Table~
\ref{table:octahedral_distortion},
most parameters such as $\lambda_{\text{oct}}$, $D$, $\sigma^2$, and $D_{\text{tilt}}$ %
are very similar in both cases. However, the angular distortion metrics $D_{\text{in}}$
and especially $D_{\text{out}}$
show significant deviation when vdW interactions are not included, leading to poorer agreement with the experimentally derived (XRD) values.

\begin{table*}[h!]
\centering
\caption{Comparison of structural parameters for (FBA)$_2$PbBr$_4$ and (R/S-FMBA)$_2$PbBr$_4$ from XRD and DFT (PBE + vdW), except one comparison to PBE alone (no vdW). Results are from this work except two XRD values from literature \cite{Zhao2023}. Values marked with $^*$ were reported in the referenced paper \cite{Zhao2023}, and we computed the others from their reported structures.}
\label{table:octahedral_distortion}
\scalebox{0.80}{%
\begin{tabular}{|c|c|c|c|c|c|c|c|c|}
\hline
\textbf{Compound} & 
\textbf{Method} &
$\lambda_{\text{oct}}\ (\times10^{-3})$ &
$D\ (\times10^{-2})$ &
$\sigma^2 \,(^\circ)^2$ &
$D_{\text{tilt}}\ (^\circ)$ &
$D_{\text{out}}\ (^\circ)$ &
$D_{\text{in}}\ (^\circ)$ &
$\Delta D_{\text{tilt}}\ (^\circ)$
\\
\hline
\multirow{2}{*}{(FBA)$_2$PbBr$_4$}
 & XRD %
   & 0.02 & 0.5 & 5.7
   & 30.2 & 0.8 & 30.2 & 0
\\
\cline{2-9}
 & DFT %
   & 0.1 & 0.9 & 7.3
   & 33.0 & 3.5 & 32.9 & 0
\\
\cline{2-9}
 & DFT, no vdW %
   & 0.1 & 0.9 & 7.3
   & 33.0 & 21.0 & 26.1 & 0
\\
\hline
\multirow{3}{*}{(R-FMBA)$_2$PbBr$_4$}
 & XRD \cite{Zhao2023}
   & 1.12$^*$ & 3.0 & 50.5$^*$
   & 37.7 & 34.3 & 16.7 & -
\\
\cline{2-9}
 & XRD %
   & 1.3 & 3.1 & 48.6
   & \shortstack{37.0 \\ 22.5}
   & \shortstack{33.8 \\ 22.0}
   & \shortstack{16.0 \\ 34.3}
   & 14.5
\\
\cline{2-9}
 & DFT %
   & 0.6 & 1.7 & 61.8
   & \shortstack{41.0 \\ 25.2}
   & \shortstack{37.7 \\ 24.9}
   & \shortstack{17.4 \\ 38.3}
   & 15.8
\\
\hline
\multirow{3}{*}{(S-FMBA)$_2$PbBr$_4$}
 & XRD \cite{Zhao2023}
   & 1.13$^*$ & 3.0 & 50.7$^*$
   & 37.7 & 34.3 & 16.7 & -
\\
\cline{2-9}
 & XRD %
   & 1.3 & 3.1 & 48.2
   & \shortstack{37.0 \\ 22.6}
   & \shortstack{36.8 \\ 22.1}
   & \shortstack{3.6  \\ 34.3}
   & 14.4
\\
\cline{2-9}
 & DFT %
   & 0.6 & 1.8 & 61.8
   & \shortstack{41.0 \\ 24.4}
   & \shortstack{37.8 \\ 24.3}
   & \shortstack{17.1 \\ 38.3}
   & 16.6
\\
\hline
\end{tabular}
}
\end{table*}

\subsection*{Optical Properties}
\begin{figure*}[ht!]
\centering

\begin{subfigure}[b]{0.32\textwidth}
    \centering
    \makebox[0pt][l]{\hspace*{-7.5em}(a)} %
    \includegraphics[width=\textwidth]{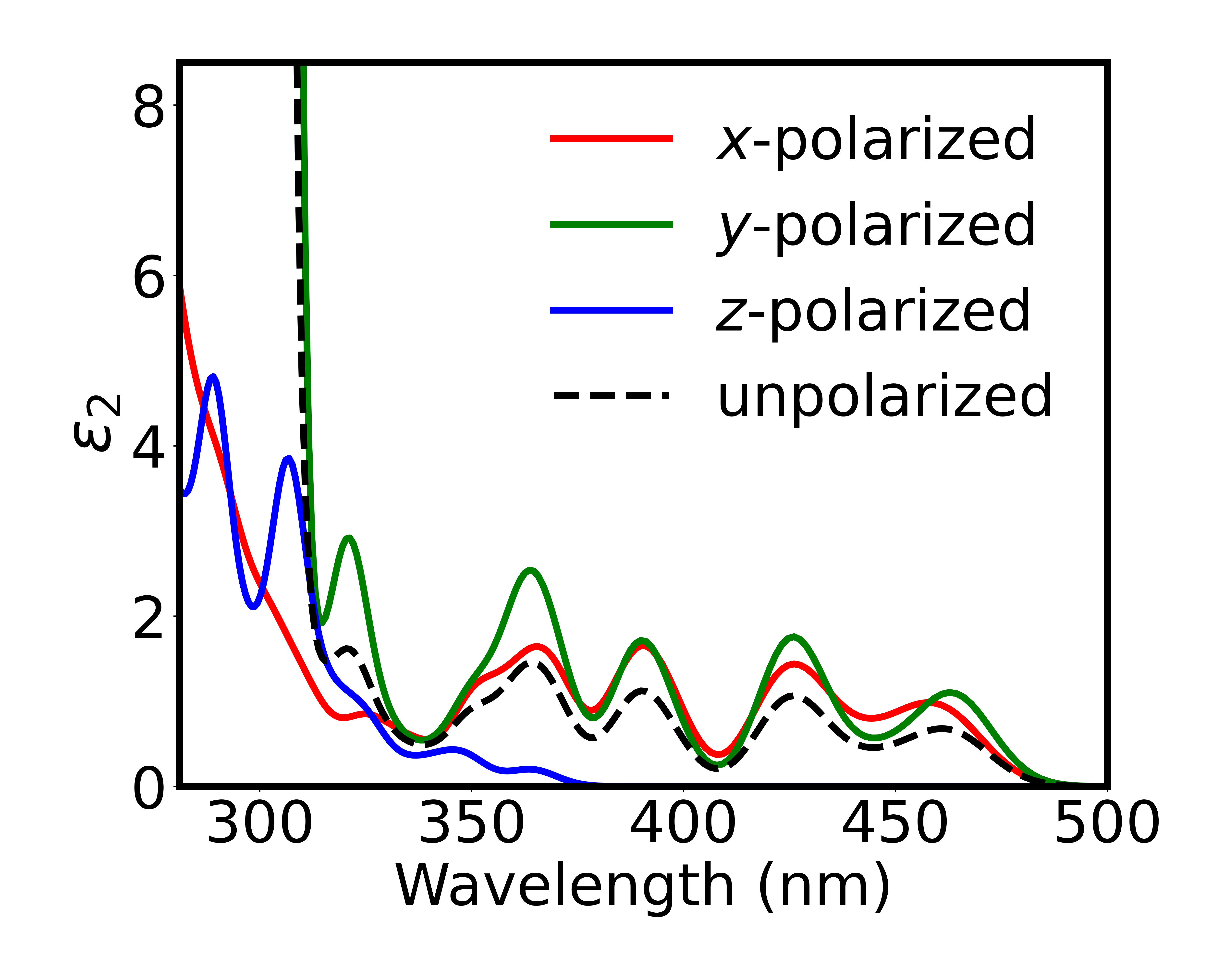}
    \label{fig:FBA_RPA_epsilon2}
\end{subfigure}
\hfill
\begin{subfigure}[b]{0.32\textwidth}
    \centering
    \makebox[0pt][l]{\hspace*{-7.5em}(b)} %
    \includegraphics[width=\textwidth]{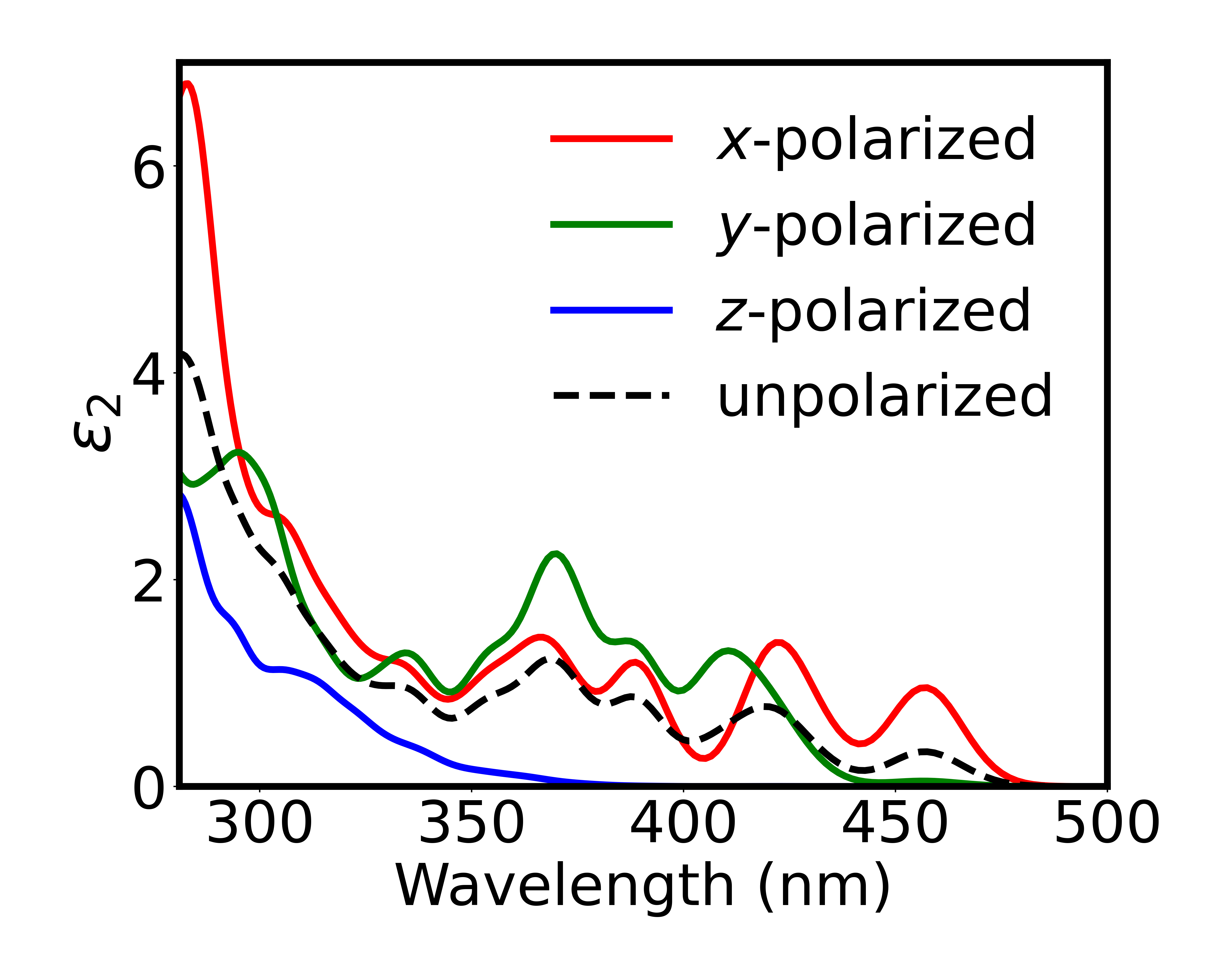}
    \label{fig:R-FMBA_RPA_epsilon2}
\end{subfigure}
\hfill
\begin{subfigure}[b]{0.32\textwidth}
    \centering
    \makebox[0pt][l]{\hspace*{-7.5em}(c)} %
    \includegraphics[width=\textwidth]{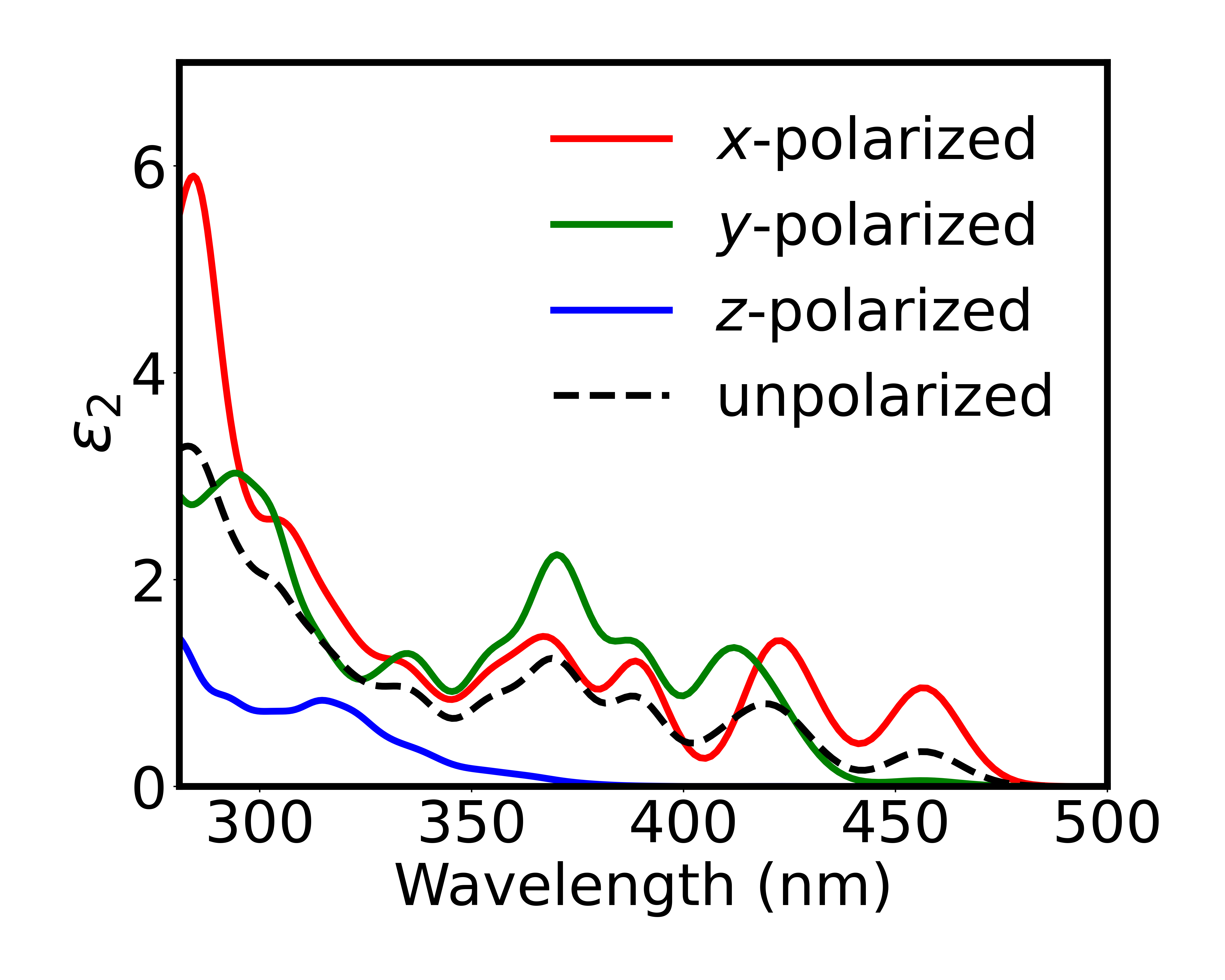}
    \label{fig:S-FMBA_RPA_epsilon2}
\end{subfigure}

\caption{Polarization-dependent optical absorption spectra, calculated with RPA, for (a) $(\mathrm{FBA})_2\mathrm{PbBr}_4$, (b) $(\mathrm{R\text{-}FMBA})_2\mathrm{PbBr}_4$, 
and (c) $(\mathrm{S\text{-}FMBA})_2\mathrm{PbBr}_4$ structures.}
\label{fig:rpa_epsilon2}
\end{figure*}

The polarized optical absorption spectra for FBA$_2$PbBr$_4$ and (R/S-FMBA)$_2$PbBr$_4$ were calculated (Fig. \ref{fig:rpa_epsilon2}). In each case, the long-wavelength onset is due to $x$- and $y$-polarized transitions, with $z$-polarized transitions contributing only below around 350 nm. Due to quantum confinement in the 2D layers, only in-plane transitions of the layer appear at low energy, and as we will see in the next section, the $z$-polarized transitions involve the organic cations. In FMBA, the onset is dominated by $y$-polarization and the $x$-polarized transitions have very low intensity at the onset.

\begin{figure*}[ht!]
\centering
\begin{tikzpicture}

    \node [anchor=north west] (imgA) at (-0.65\linewidth, 0.80\linewidth) {\includegraphics[scale=0.33]{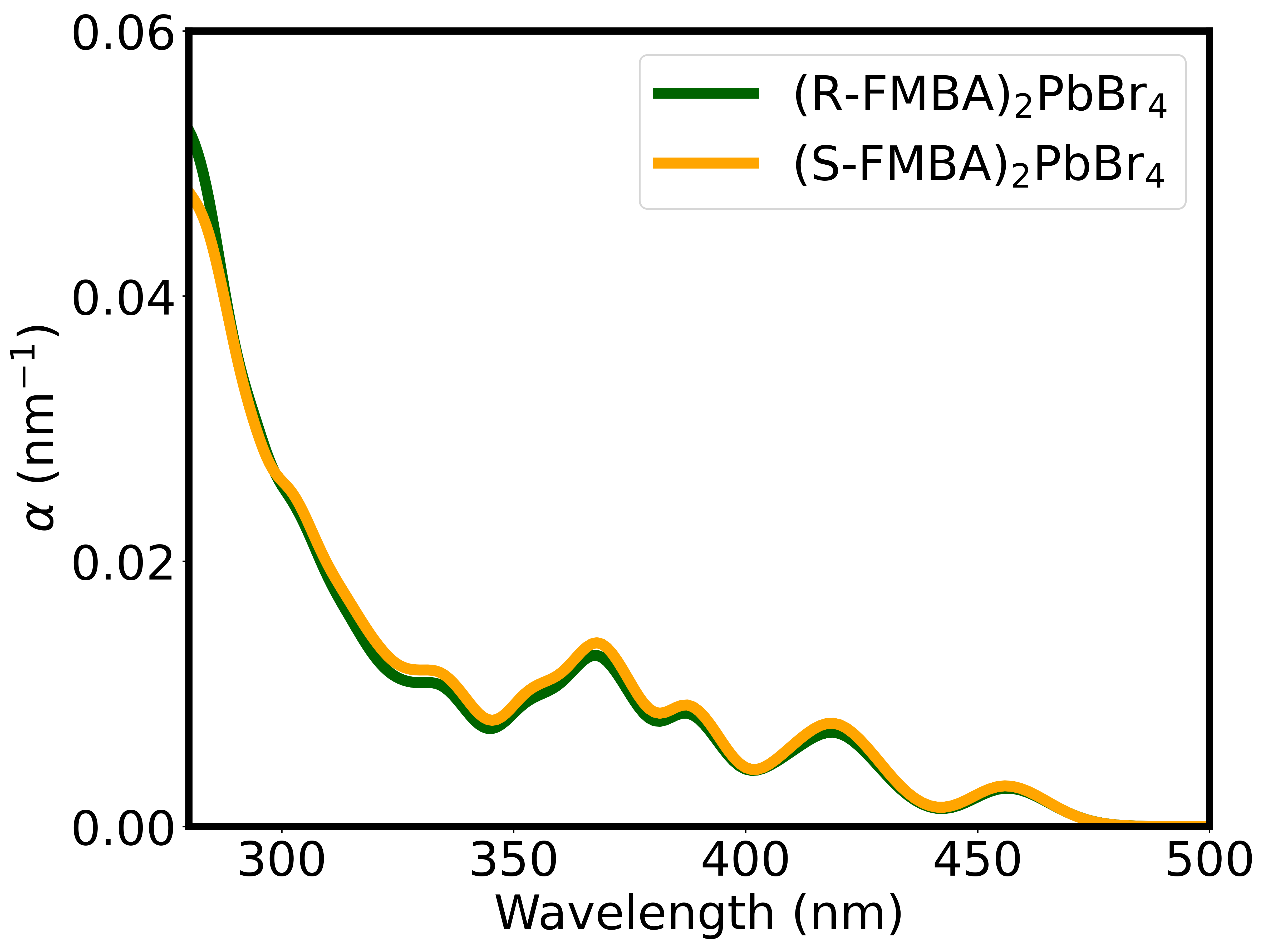}
    };
    \draw [anchor=north west] (-0.63\linewidth, 0.84\linewidth) node {\textbf{(a)}};

    \node [anchor=north west] (imgA) at (-0.15\linewidth, 0.80\linewidth) {\includegraphics[scale=0.33]{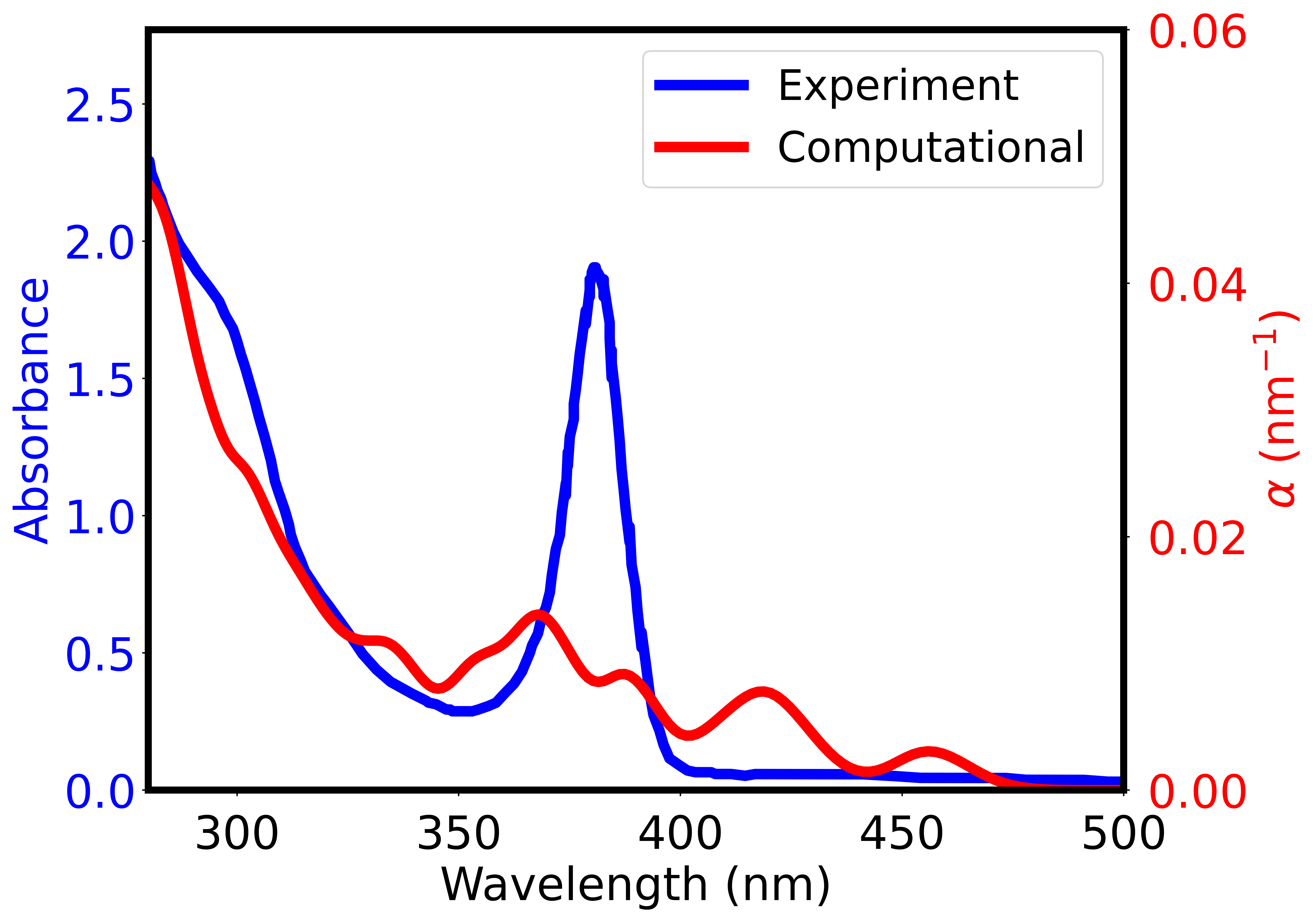}
    };
    \draw [anchor=north west] (-0.14\linewidth, 0.83\linewidth) node {\textbf{(b)}};

\end{tikzpicture}
\caption{(a) Computed absorption coefficients for (R/S-FMBA)$_2$PbBr$_4$, and (b) computed and measured absorption spectra for (S-FMBA)$_2$PbBr$_4$.}
\label{fig:optical_spectra_comparison}
\end{figure*}

Comparing R and S structures of FMBA, only slight differences are found in the absorption coefficients (Fig. \ref{fig:optical_spectra_comparison}(a)). We compared to the experimental spectrum in Fig. \ref{fig:optical_spectra_comparison}). 
The experimental spectrum displays a prominent peak at 381 nm, comparable to the peak around 375 nm previously reported \cite{Zhao2023}, and a large rise around 300 nm.
By contrast, the computed spectra show an onset at a longer wavelength around 475 nm with several small peaks, but do also have the large rise around 300 nm. The disagreement in the onset and lack of well defined peak is because the DFT-RPA level of theory used in these calculations cannot describe excitonic peaks \cite{Deslippe_2012, lee2023one,mcclintock2023surface}, as well as the lack of quasiparticle renormalization and neglect of spin-orbit coupling.

\subsection*{Partial Density of States (PDOS)}
\begin{figure*}[ht!]
\centering

\begin{subfigure}[b]{0.32\textwidth}
    \centering
    \makebox[0pt][l]{\hspace*{-7.5em}(a)} %
    \includegraphics[width=\textwidth]{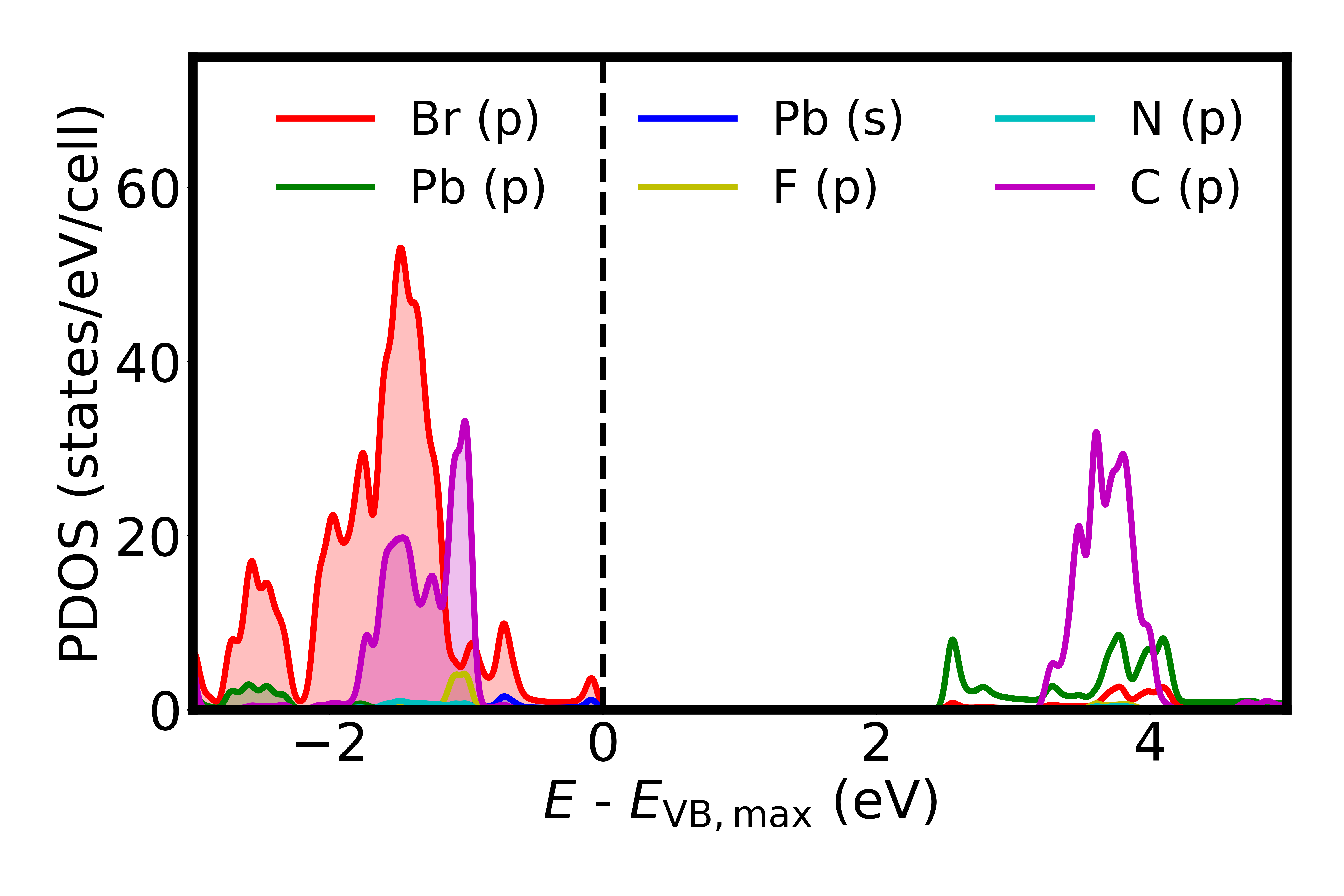}
    \label{fig:FBA_PDOS}
\end{subfigure}
\hfill
\begin{subfigure}[b]{0.32\textwidth}
    \centering
    \makebox[0pt][l]{\hspace*{-7.5em}(b)} %
    \includegraphics[width=\textwidth]{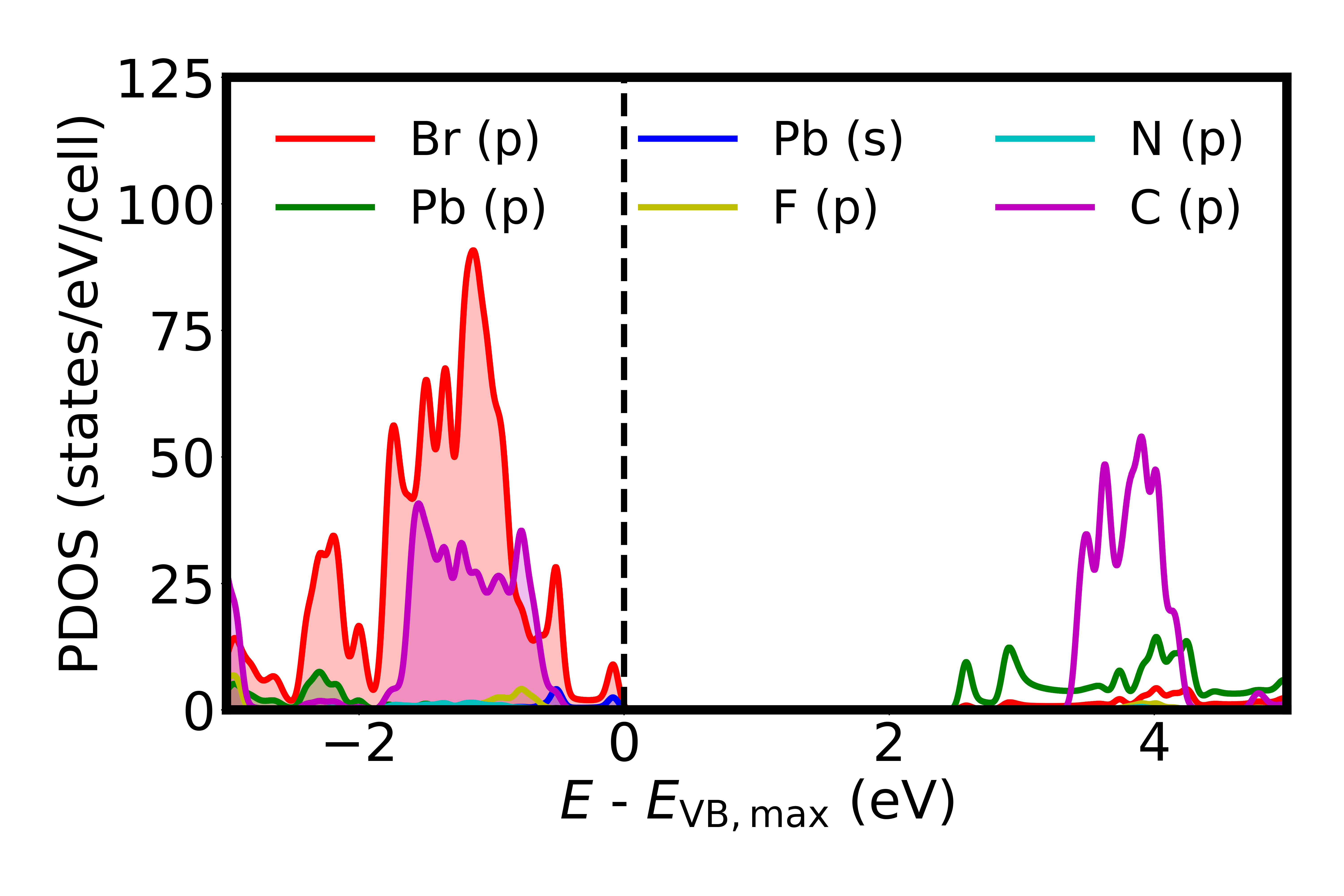}
    \label{fig:R-FMBA_PDOS}
\end{subfigure}
\hfill
\begin{subfigure}[b]{0.32\textwidth}
    \centering
    \makebox[0pt][l]{\hspace*{-7.5em}(c)} %
    \includegraphics[width=\textwidth]{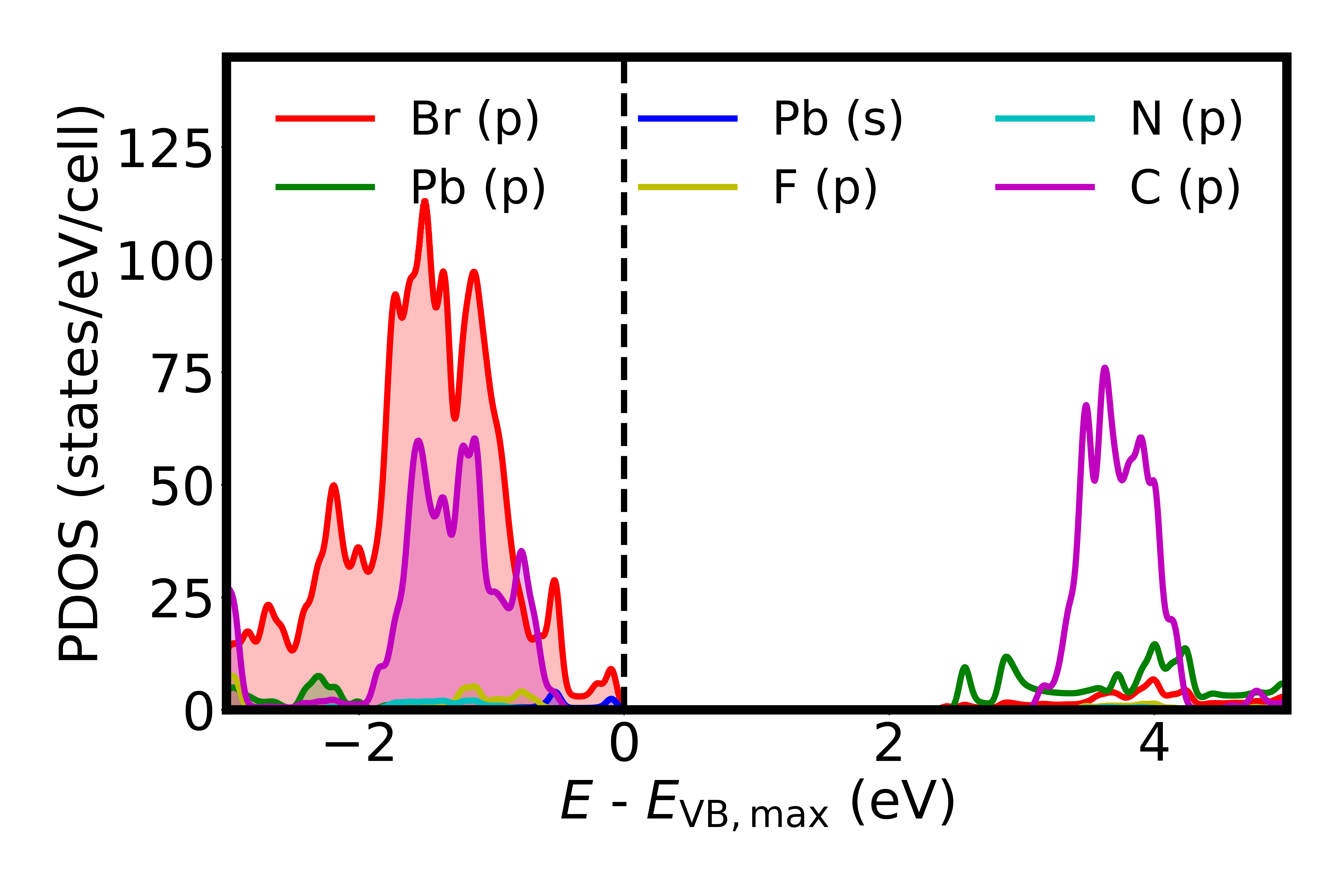}
    \label{fig:S-FMBA_PDOS}
\end{subfigure}

\caption{Partial density of states (PDOS) showing major orbital contributions near valence and conduction band edges in (a) $(\mathrm{FBA})_2\mathrm{PbBr}_4$, (b) $(\mathrm{R\text{-}FMBA})_2\mathrm{PbBr}_4$, 
and (c) $(\mathrm{S\text{-}FMBA})_2\mathrm{PbBr}_4$ structures.}
\label{fig:pdos_all}
\end{figure*}

PDOS analysis was performed to assess which orbitals contributed the most to the transitions (Fig. \ref{fig:pdos_all}.
PDOS analysis for both (FBA)$_2$PbBr$_4$ and (R-FMBA)$_2$PbBr$_4$ showed that the Br p-orbital contribution near the valence bands was dominant compared to all other orbitals. The second highest contribution near the valence bands was from the Pb s-orbital. The Pb p-orbital contribution was prominent in the conduction bands closer to the valence bands.

\subsection*{Electronic Structure}
\begin{figure*}[ht!]
\centering

\begin{subfigure}[b]{0.32\textwidth}
    \centering
    \makebox[0pt][l]{\hspace*{-7.5em}(a)} %
    \includegraphics[width=\textwidth]{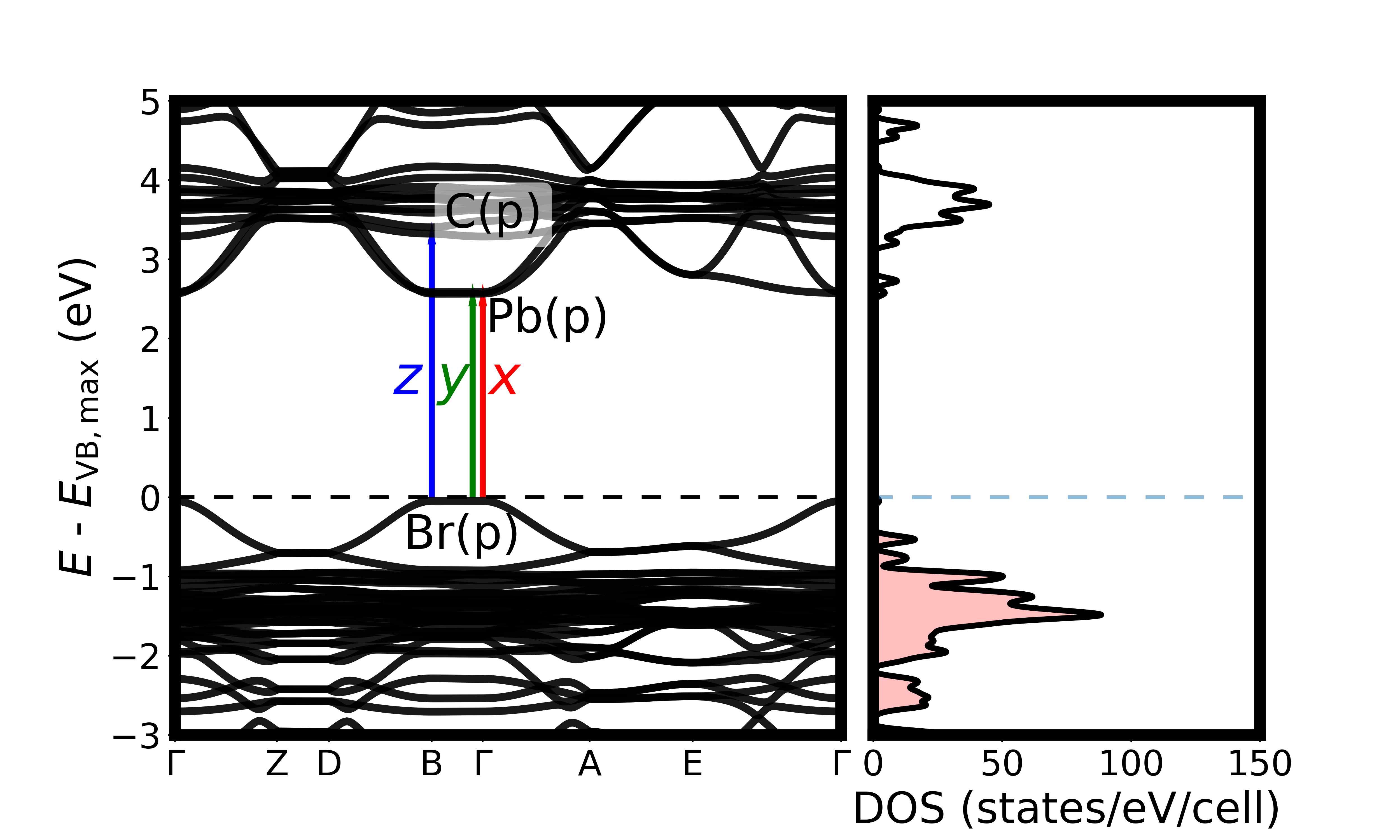}
    \label{fig:FBA_BS_DOS}
\end{subfigure}
\hfill
\begin{subfigure}[b]{0.32\textwidth}
    \centering
    \makebox[0pt][l]{\hspace*{-7.5em}(b)} %
    \includegraphics[width=\textwidth]{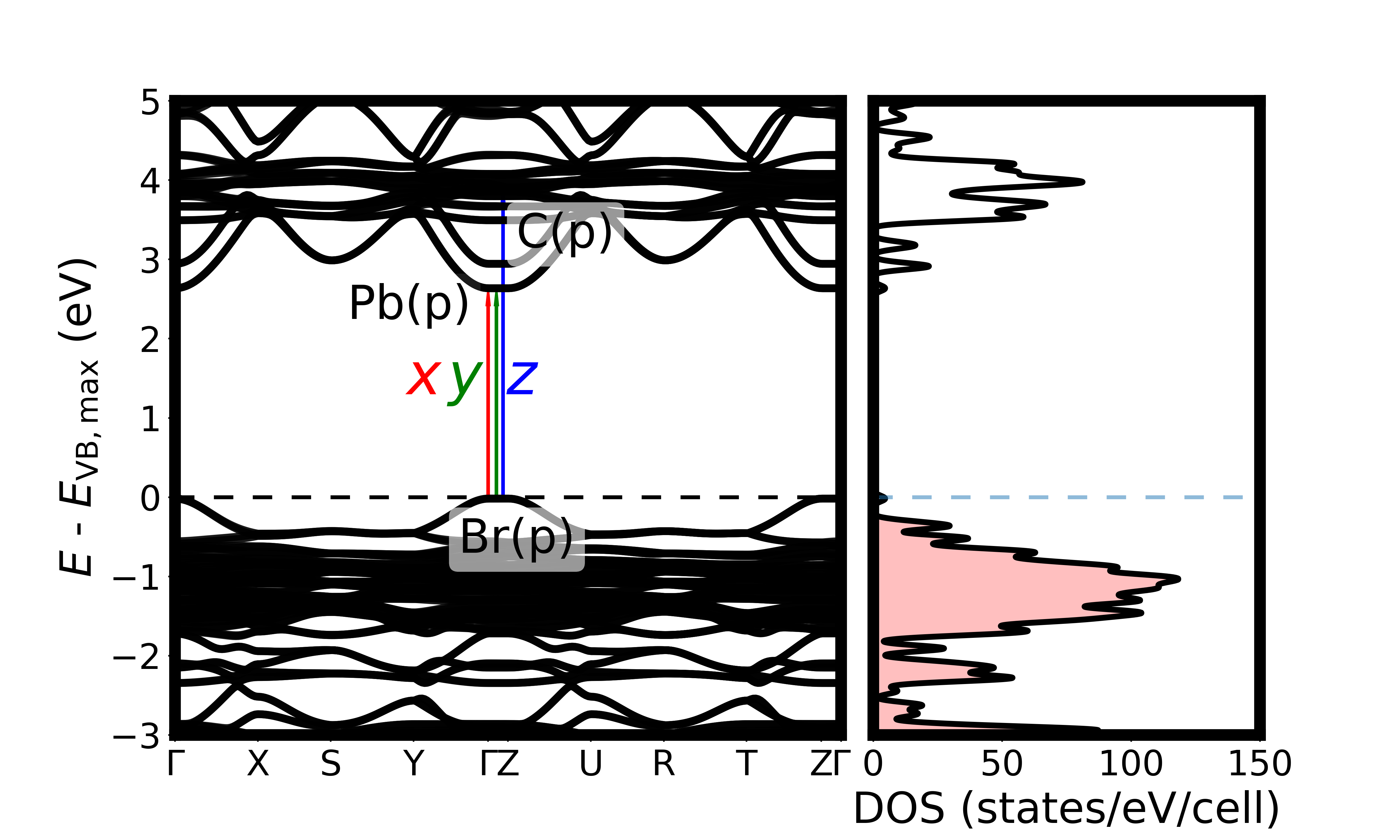}
    \label{fig:R-FMBA_BS_DOS}
\end{subfigure}
\hfill
\begin{subfigure}[b]{0.32\textwidth}
    \centering
    \makebox[0pt][l]{\hspace*{-7.5em}(c)} %
    \includegraphics[width=\textwidth]{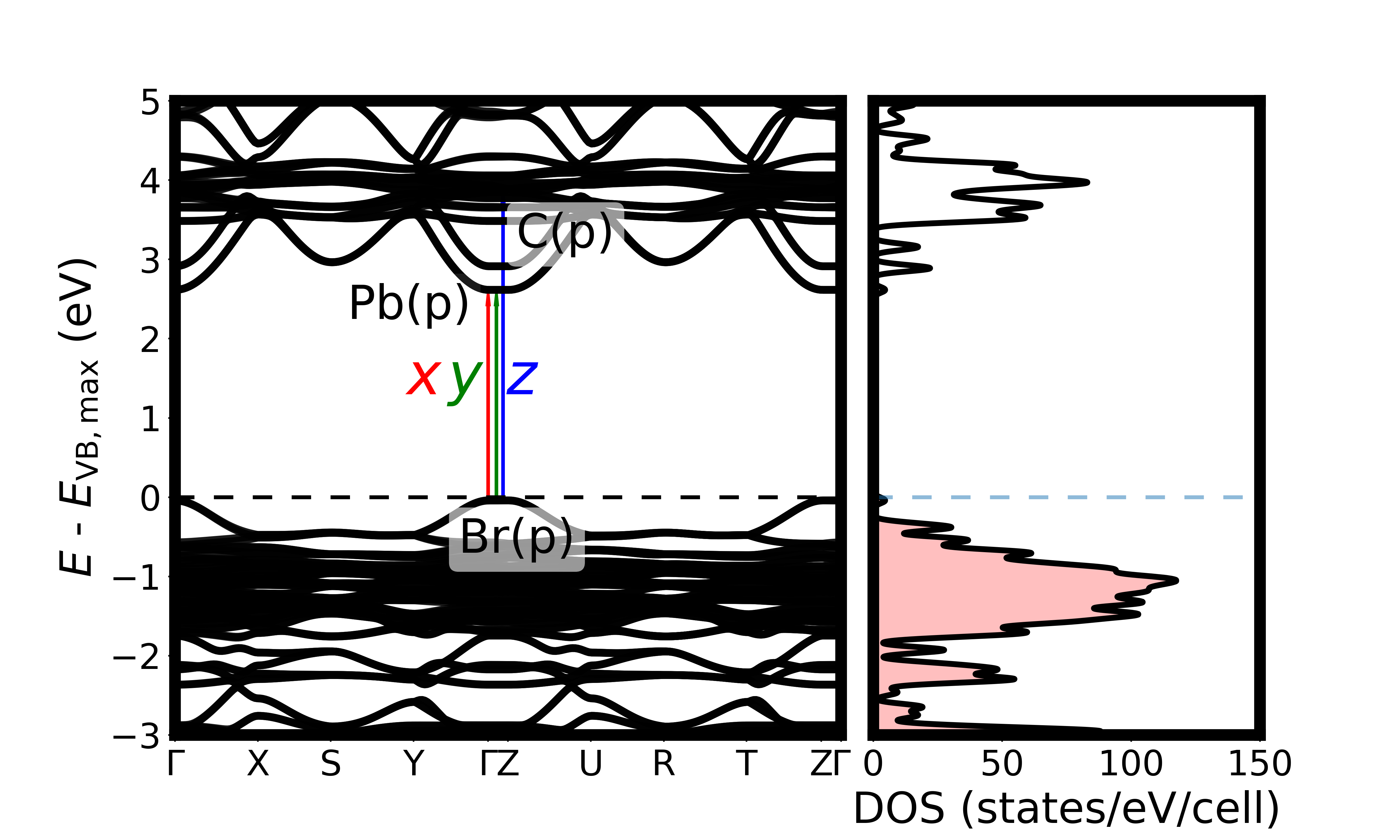}
    \label{fig:S-FMBA_BS_DOS}
\end{subfigure}

\caption{Calculated electronic band structures and corresponding density of states for (a) $(\mathrm{FBA})_2\mathrm{PbBr}_4$, (b) $(\mathrm{R\text{-}FMBA})_2\mathrm{PbBr}_4$, and (c) $(\mathrm{S\text{-}FMBA})_2\mathrm{PbBr}_4$. The lowest-energy optical transitions are marked for $x$, $y$, and $z$ polarizations, and the predominant orbital character is marked for the valence and conduction bands involved in these levels.}
\label{fig:bs_dos_stacked}
\end{figure*}

The electronic bandstructure and significant orbital contributions to the electronic transitions were calculated for each material, and are shown in Fig. \ref{fig:bs_dos_stacked}. A direct bandgap was found in each case.
The band gap of (FBA)$_2$PbBr$_4$ was calculated to be 2.54 eV. The lowest-energy transitions were due to $x$ and $y$-polarizations, with $y$ somewhat more intense.
The transitions near the band gap primarily involved orbitals near the $\Gamma$ and B points in the Brillouin zone. Br p and Pb p orbitals had major contributions to The $x$- and $y$-polarized transitions were predominantly from Br p orbitals to Pb p orbitals, whereas $z$-polarized transitions were predominantly from Br p to C p.
(R/S-FMBA)$_2$PbBr$_4$ had a slightly larger calculated band gap of 2.65 eV. The lowest-energy transitions similarly were due to $x$ and $y$-polarizations near the $\Gamma$ point, though with very low intensity for $x$ near the onset, and the main orbital contributions to the transitions are the same as in FBA. 
A summary of the electronic bandgaps and the electronic transitions, from valence bands (VB) to conduction bands (CB), for different polarizations are shown in table \ref{table:transition_bands}. %

\begin{table*}[h!]
\centering
\caption{Lowest-energy optical transitions for each polarization, with corresponding conduction band (CB) and valence band (VB) indices (counting starts with 1 at the band edge) and their principal atomic orbital contributions, and predominant $k$-point in the bandstructure, for 
(FBA)\textsubscript{2}PbBr\textsubscript{4} and (R/S-FMBA)\textsubscript{2}PbBr\textsubscript{4}.}
\label{table:transition_bands}
\begin{tabular}{|c|c|c|c|c|c|c|c|}
\hline
\textbf{Material} 
& \boldmath$E_g$ (\textbf{eV}) 
& \textbf{Polarization} 
& \textbf{CB}
& \textbf{CB orb.} 
& \textbf{VB} 
& \textbf{VB orb.} 
& \textbf{$k$-Point} \\ 
\hline
\multirow{3}{*}{(FBA)\textsubscript{2}PbBr\textsubscript{4}}
  & 2.54 & $x$ & 1 & Pb p & 1 & Br p & $\Gamma$ \\
  & 2.54 & $y$ & 1 & Pb p & 1 & Br p & $\Gamma$ \\
  & 3.30 & $z$ & 3 & C p  & 1 & Br p & B \\
\hline
\multirow{3}{*}{(R-FMBA)\textsubscript{2}PbBr\textsubscript{4}}
  & 2.65 & $x$ & 1 & Pb p & 2 & Br p & $\Gamma$ \\
  & 2.65 & $y$ & 1 & Pb p & 1 & Br p & $\Gamma$ \\
  & 3.79 & $z$ & 10 & C p & 1 & Br p & $\Gamma$ \\ 
\hline
\multirow{3}{*}{(S-FMBA)\textsubscript{2}PbBr\textsubscript{4}}
  & 2.65 & $x$ & 1 & Pb p & 2 & Br p & $\Gamma$ \\
  & 2.65 & $y$ & 1 & Pb p & 1 & Br p & $\Gamma$ \\
  & 3.75 & $z$ & 10 & C p & 1 & Br p & $\Gamma$ \\ 
\hline
\end{tabular}
\end{table*}

\section{Conclusion}
We employed density functional theory to investigate the structural and octahedral distortion parameters of 2D chiral hybrid perovskites, specifically (FBA)$_2$PbBr$_4$ and (R/S-FMBA)$_2$PbBr$_4$, in conjunction with experimental synthesis and characterization with XRD and UV/Vis spectroscopy. The calculations were performed with and without van der Waals corrections and were systematically compared with experimental data obtained from XRD. 
The results demonstrate that van der Waals corrections improve the accuracy of DFT-PBE structures for these materials, with reduced discrepancies in lattice parameters and octahedral distortions when compared to XRD, although tilt angles remain somewhat overestimated. 
Key distortion parameters within octahedra as well as tilt angles between octahedra ($D_{\text{tilt}}$, $D_{\text{out}}$, and $D_{\text{in}}$), were analyzed. We find some discrepancies between our XRD and previous XRD results \cite{Zhao2023}, indicating the sensitivity of these quantities. We found a very large value of the difference $D_{\text{tilt}}$ for FMBA, which is associated with chiral symmetry-breaking and helps to explain the strongly circularly polarized emission that has been observed \cite{Zhao2023}, and indicates potential for large spin splittings in these materials for spintronic applications \cite{Jana}.

Calculated polarized optical spectra showed that low-energy transitions are polarized in plane, and are Br p to Pb p predominantly, whereas at higher energies $z$-polarized Br p to C p transitions are found. There is a significant excitonic peak in experiment which was not captured in our RPA calculations.

We provide a Python code that we developed to facilitate calculations of octahedral distortion parameters in perovskites, to enable investigations of octahedral distortions in broader classes of  perovskites and advance our understanding of structure-property relations in chiral and layered organic-inorganic perovskites.

\section*{Acknowledgments}
Computational work by M.M.M. and D.A.S. was supported by the U.S. Department of Energy, National Nuclear Security Administration, Minority Serving Institution Partnership Program (DE-NA0003984). Experimental work by J.V., A.B.-A., and B.M. was supported by the National Science Foundation (NSF) (DMR-2204466).
This research used resources of the National Energy Research Scientific Computing Center; a DOE Office of Science User Facility supported by the Office of Science of the U.S. Department of Energy under Contract No. DE-AC02-05CH11231 using NERSC award FES-ERCAP0028527.

\bibliographystyle{wiley-chemistry} %
\bibliography{example_refs} 

\clearpage

\noindent\rule{11cm}{2pt}
\begin{minipage}{11cm}
\includegraphics[width=11cm]{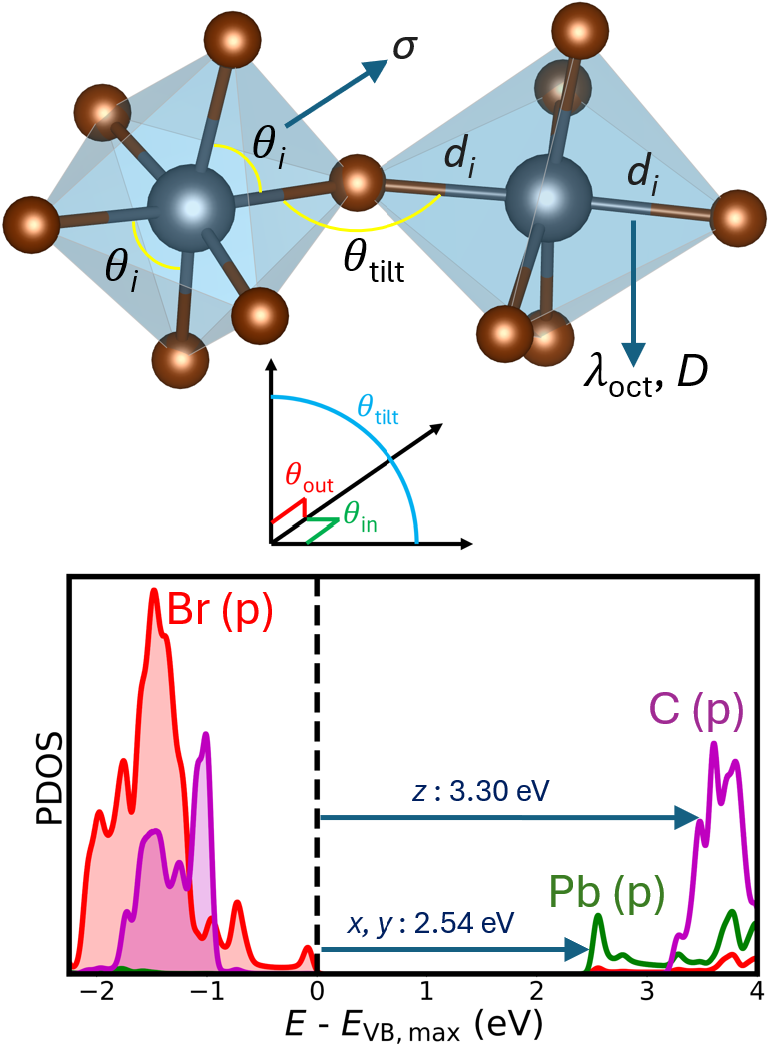}
\end{minipage}
\begin{minipage}{11cm}
\large\textsf{Octahedral distortion parameters in 2D hybrid organic-inorganic perovskites, as calculated by our Python code, are correlated with electronic and optical properties. We computationally analyze structure, orbitals, and anisotropic optical transitions of synthesized chiral and achiral lead bromide perovskites.
}
\end{minipage}
\noindent\rule{11cm}{2pt}

\end{document}